\documentclass[11pt]{article}
\pdfoutput=1

\usepackage[a4paper,text={16.8cm,22.4cm}]{geometry}
\usepackage{amsmath,amsfonts,slashed,amssymb,tikz,bm,psfrag,graphicx,color,dsfont,euscript}
\usepackage{multicol}
\usepackage{float}
\usepackage{empheq}
\RequirePackage[sort&compress,square,comma,numbers]{natbib}
\allowdisplaybreaks
\renewcommand{\arraystretch}{1.2}

\newcommand{\F}{{\EuScript F}}
\newcommand{\G}{{\EuScript G}}
\newcommand{\J}{{\EuScript J}}

\begin{document}

\begin{titlepage}
\vfill
\begin{center}
{\Large \bf Revisiting radiative leptonic $B$ decay}\\[0.7cm]
{\bf $^{a}$Yue-Long Shen\footnote{{ email address: shenylmeteor@ouc.edu.cn}}, $^{b}$Yan-Bing Wei\footnote{{ email address: weiyb@nankai.edu.cn}}, $^{b}$Xue-Chen Zhao\footnote{{ email address: zxc@mail.nankai.edu.cn}}, $^{c}$Si-Hong Zhou\footnote{{ email address: shzhou@imu.edu.cn}}}\\[0.5cm]
{\it $^a$ College of Information Science and Engineering, Ocean
University of China, Qingdao 266100, P.R. China }\\
{\it $^b$School of Physics, Nankai University, Weijin Road 94, 300071 Tianjin, P.R. China}\\
{\it $^c$School of Physical Science and Technology,
Inner Mongolia University, Hohhot 010021,P.R. China}
\end{center}
\vfill
\begin{abstract}
 In this paper, we summarize the existing methods of solving the  evolution equation of
the leading-twist $B$-meson LCDA.
Then, in the Mellin space, we derive a factorization formula with next-to-leading-logarithmic (NLL) resummation for the form factors $F_{A,V}$ in the $B \to \gamma \ell\nu$
decay at leading power in $\Lambda/m_b$. Furthermore, we investigate the power suppressed local contributions, factorizable non-local contributions (which are suppressed by $1/E_\gamma$ and $1/m_b$), and soft contributions to the form factors.  In the numerical analysis, which employs the two-loop-level hard function and the jet function, we find that both the resummation effect and the power corrections can sizably decrease the form factors.
Finally, the integrated branching ratios are also calculated for comparison with future experimental data. \end{abstract}
\vfill
\end{titlepage}

\section{Introduction}

The radiative leptonic decay of the $B$ meson is of interest as it is the most important channel
to extract the parameters of the $B$-meson light-cone distribution amplitudes (LCDAs) and to test the factorization theorem when
the emitted photon is energetic. A precision study of this mode can also be helpful in decreasing the  background to the purely leptonic decay process $B^- \to \ell^-\nu$, which is important for determining the CKM matrix element $V_{ub}$. The radiative leptonic $B \to \gamma \ell \nu$ decay amplitude is defined by the
QCD matrix element
\begin{eqnarray}
{\cal A}(B \to \gamma \, \ell \, \nu )
=\frac{G_F \, V_{ub}} {\sqrt{2}} \, \left \langle \gamma(p) \, \ell(p_{\ell}) \, \nu(p_{\nu}) \left |
\left [ \bar{\ell} \, \gamma_{\mu} \, (1- \gamma_5) \, \nu  \right ] \,\,
\left [ \bar u \, \gamma^{\mu} \, (1- \gamma_5) \, b \right ] \right | B^{-}(p_B) \right \rangle  \,.
\label{def: full decay amplitude}
\end{eqnarray}
In the  rest frame of the $B$ meson with momentum
$p_B = m_B \, v$ , it is convenient to introduce two light-cone vectors $n_{\mu}$ and $\bar{n}_{\mu}$ with the definitions
\begin{eqnarray}
p_{\mu}=\frac{n \cdot p}{2}\, \bar{n}_{\mu} \equiv E_{\gamma} \, \bar{n}_{\mu}\,,
\qquad v_{\mu} =\frac{ n_{\mu} + \bar{n}_{\mu}} {2} \,.
\end{eqnarray}
At leading order in QED and considering the constraints from the Ward identity, the amplitude can be parameterized as \cite{Beneke:2011nf,Wang:2016qii}
\begin{eqnarray}
{\cal A}(B \to \gamma \, \ell \ \nu)
 \rightarrow  {G_F \, V_{ub} \over \sqrt{2}} \, \left ( i \, g_{em} \, \epsilon_{\nu}^{\ast}  \right ) \,
v \cdot p \, \bigg \{ - i  \, \epsilon_{\mu \nu \rho \sigma}
\, n^{\rho} \, v^{\sigma} \, F_V(E_\gamma)
+ g_{\mu \nu} \, \hat F_A(E_\gamma)  \bigg \} \,,
\label{B to photon form factor}
\end{eqnarray}
where the contribution from  final-state radiation
is accounted for by the redefinition of the axial form factor $\hat F_A(E_\gamma)$.

At leading power in $\Lambda/m_b$, the QCD factorization formula has been derived for the $B \to \gamma$
form factors $F_{A,V}$  \cite{Korchemsky:1999qb,DescotesGenon:2002mw} and was confirmed under the framework of  soft-collinear effective theory (SCET) \cite{Lunghi:2002ju,Bosch:2003fc}.
The form factors $F_{A,V}(E_\gamma)$ can be factorized into a convolution of the hard function, jet function, and $B$-meson LCDA. The hard function arises from the matching between  heavy-to-light current in the QCD and SCET${\rm _I}$ operators, and it has been calculated up to the two-loop level \cite{Beneke:2011nf}. The jet function can be obtained from the matching between SCET$_{\rm I}$ and SCET$_{\rm II}$ \cite{Lunghi:2002ju}, and the next-to-next-to-leading-order (NNLO) correction has been recently obtained \cite{Liu:2020ydl}. The matrix elements of the SCET$_{\rm II}$ operators are actually the definition of $B$-meson LCDA. All the ingredients in the factorization formula depend on the factorization scale, and the radiative corrections lead to large logarithmic terms, which need to be resummed. For the hard function, the three-loop anomalous dimension is known \cite{Henn:2019swt,Chetyrkin:2003vi,Becher:2009qa,Bruser:2019yjk}, and the two-loop-level anomalous dimensions both for the $B$-meson LCDA and the jet function have recently been calculated \cite{Braun:2019wyx,Liu:2020ydl}.
Therefore, the sufficient condition for a complete NLL resummation is readily available. It was first derived in \cite{Galda:2020epp} by performing a Laplace transformation of the $B$-meson LCDA.

Although the factorization formula of $B \to \gamma  \ell\nu$ decay is well established at leading power, the  power corrections are important for finite bottom-quark mass. The subleading-power corrections suppressed by a factor of ${\cal O}(\Lambda/ m_b)$ were considered  at tree level \cite{Beneke:2011nf}, where a
symmetry-preserving form factor $\xi(E_{\gamma})$ was introduced
to parameterize the non-local power corrections.  The  soft contribution from the endpoint region of the momentum of the light quark inside the $B$ meson was first studied using dispersion relation and quark-hadron duality in \cite{Braun:2012kp}. The QCD correction to the soft contribution at one loop and the contribution from
three-particle LCDAs were computed in
\cite{Wang:2016qii}. In a comprehensive study on the local and non-local power suppressed contributions, the soft contribution and the higher-twist contribution to the $B \to \gamma  \ell \nu$ decay  were presented \cite{weiyb}. The contribution from the hadronic structure of the photon, which can be defined by the matrix elements of power-suppressed SCET operators,
was studied in \cite{Ball:2003fq, Wang:2018wfj}. Morevoer, based on  transverse-momentum-dependent factorization, the power corrections to $B \to \gamma \ell\nu $ decay were investigated in \cite{Shen:2018abs}. All these studies indicate that the power-suppressed contribution is sizable and should not be neglected in the determination of the first inverse moment of the $B$-meson LCDA.

In this paper, we make improvements from two aspects. The first is to derive the scale-independent leading-power factorization formula at the NLL level in the Mellin space, and the second is to perform the phenomenological analysis after combining the NLL leading-power result with the power corrections.
This paper is organized as follows. In the next section, we
review the evolution of the leading-twist $B$-meson LCDA. In the third section, we derive the scale-independent factorization formula of the $B\to \gamma$ form factors and discuss the power-suppressed contributions; this is followed by the phenomenological  analysis. Concluding discussions are presented in the final section.

\section{The evolution of the $B$-meson LCDA}
The $B$-meson LCDA is one of the most important ingredients of the QCD factorization formula for exclusive $B$ decays.
The two-particle LCDAs of the $B$ meson in the heavy-quark effective theory (HQET)  can be obtained from the coordinate-space matrix elements \cite{Grozin:1996pq}
\begin{eqnarray}\langle0|\bar q^\beta(z)[z,0]h_v^\alpha(0)|\bar B(v)\rangle=-{i\tilde f_B m_B\over 4}
\left[{1+\not\!v\over 2}\left\{\tilde\Phi^+{(t,z^2)}+{\tilde\Phi^-{(t,z^2)}-\tilde\Phi^+{(t,z^2)}\over t}\not\!z\right\}\right]^{\alpha\beta}\, .
\end{eqnarray}
The LCDAs $\tilde\Phi^\pm{(t,z^2)}$ (in curly brackets) can be expanded around $z^2= 0$. In the limit $z^2\to 0$, $t\to \tau={ n\cdot z/2}$, the $B$-meson LCDAs in the momentum space are defined through the Fourier transformation
\begin{eqnarray}
\phi^{\pm}(\omega)=\int {d\tau\over 2\pi} \, e^{i\omega \tau} \, \tilde\Phi^{\pm}({\tau})\, .
\end{eqnarray}
At leading power, only $\phi^{+}(\omega)$ is relevant in the $B \to \gamma \ell\nu $ decay, and the evolution equation of $\phi^{+}(\omega)$ is the well-known Lange-Neubert equation \cite{Lange:2003ff}:
\begin{eqnarray}
\frac { d } { d \ln \mu } \phi  ^ { + } ( \omega ,\mu ) &=&
-  \int _ { 0 } ^ {
\infty } d \omega ^ { \prime }\Gamma_+(\omega,\omega',\mu)\phi^{+} \left( \omega ^ {
\prime } , \mu \right),\nonumber \\
\Gamma_+(\omega,\omega',\mu) &=& \left( \Gamma_{\rm cusp} \ln
\frac { \mu} { \omega } + \gamma \right) \delta \left( \omega -
\omega ^ { \prime } \right) + \omega \, \Gamma_{\rm cusp}\,\Gamma(\omega,\omega')\, ,
\label{LN:evolution:eq}
\end{eqnarray}
where $\mu$ is the renormalization scale. At the one-loop level, the
anomalous dimensions are
\begin{align}
\Gamma_{\rm cusp}=&~\sum_{n=1}\Gamma_{\rm cusp}^{(n)}\,
\Big(\frac{\alpha_s}{4\pi}\Big)^n,
&\gamma=&~\sum_{n=1}\gamma^{(n)}\,
\Big(\frac{\alpha_s}{4\pi}\Big)^n,
\nonumber \\
\Gamma_{\rm cusp}^{(1)}= &~4\,C_F,
&\gamma^{(1)}=&-2\,C_F,\nonumber \\
\Gamma(\omega,\omega') =&- \left[ \frac { \theta \left( \omega ^
{ \prime } - \omega \right) } { \omega^{ \prime } \left( \omega ^
{ \prime } - \omega \right) }+\frac { \theta \left( \omega -
\omega ^ { \prime } \right) } { \omega \left( \omega - \omega ^ {
\prime } \right) } \right] _ { + }\, ,&&
\end{align}
with the ``plus" function defined as
\begin{eqnarray}
\int^\infty_0 dy \,\Big [ f(x,y) \Big ]_{+} g(y)=
\int^\infty_0 dy \, f(x,y) \,\Big[ g(y)-g(x) \Big]\,.
\end{eqnarray}
In the position space, the evolution equation of the $B$-meson LCDA takes the form \cite{Kawamura:2010tj}
\begin{eqnarray}
\frac { d } { d \ln \mu } \tilde\Phi  ^ { + } ( t ,\mu ) &=&
- \big[\Gamma_{\rm cusp}(\alpha_s)\ln{it\tilde\mu}+\gamma_+(\alpha_s)-\gamma_F(\alpha_s)\big]\,\tilde\Phi  ^ { + } ( t ,\mu )+\int _ { 0 } ^ {
1} d z \,K(z,\alpha_s)\,\tilde\Phi  ^ { + } ( zt ,\mu ) \, ,
\label{positionspace}
\end{eqnarray}
where the at one-loop level,
\begin{align}
 \tilde \mu=\mu \, e^{\gamma_E}, \qquad \gamma_+({\alpha_s})=-{\alpha_sC_F\over 4\pi} , \qquad
\gamma_F({\alpha_s})=-{3\alpha_sC_F\over 4\pi},
\qquad K(z,\alpha)=\frac{\alpha_s\,C_F}{\pi}\left({z\over 1-z}\right)_+\, .
\end{align}
Whether in the momentum space or in the position space, the evolution equation of the $B$-meson LCDA is the integro-differential equation. It is difficult to solve directly: it must be simplified it by an integral transformation. To date, there exist the following treatments:
\begin{itemize}
\item Performing the Fourier transformation with respect to $\ln (\omega/\mu)$ (or the Mellin transform  $\langle\omega^{N-1}\rangle$ for $N=i\theta$):
\begin{eqnarray}
\varphi^+(\theta,\mu)=\int_0^\infty {d\omega \over \omega}\phi^+(\omega)\left({\omega\over \mu}\right)^{-i\theta}\, .
\end{eqnarray}
Then, the evolution kernel of $\varphi^+(\theta,\mu)$ is obtained as \cite{Lee:2005gza}
\begin{eqnarray}
\varphi_B^+(\theta,\mu)=e^{V(\mu,\mu_0)-2\gamma_E g}\left({\mu\over \mu_0}\right)^{i\theta}{
\Gamma(1-i\theta)\Gamma(1+i\theta-g)\over \Gamma(1+i\theta)\Gamma(1-i\theta+g)}\varphi_B^+(\theta+ig,\mu_0)\, .
\end{eqnarray}
Through the inverse Fourier transformation, we arrive at the solution to the evolution equation in the momentum space: 
 \begin{align}
 \phi^+(\omega,\mu)=e^{V(\mu,\mu_0)-2\gamma_E g}{\Gamma(2-g)\over \Gamma(g)}\int_0^\infty{d\omega'\over \omega'}\phi^+(\omega',\mu_0)\left({\omega_{>}\over \mu_0}\right)^g{\omega_{<}\over \omega_{>}}{}_2F_1
 \left(1-g,2-g;2;{\omega_{<}\over \omega_{>}}\right)\, ,
 \end{align}
 where $\omega_<=\min(\omega,\omega'),\omega_>=\max(\omega,\omega')$,
and the functions $V$ and $g$ take the form
\begin{eqnarray}
V\left(\mu, \mu_{0}\right)
&=&-\int_{\alpha_{s}\left(\mu_{0}\right)}^{\alpha_{s}(\mu)} \frac{d \alpha}{\beta(\alpha)}
\left[\Gamma_{\mathrm{cusp}}(\alpha) \int_{\alpha_{s}\left(\mu_{0}\right)}^{\alpha}
\frac{d \alpha^{\prime}}{\beta\left(\alpha^{\prime}\right)}+\gamma(\alpha)\right]\, ,
 \nonumber\\
g &\equiv& g\left(\mu, \mu_{0}\right)
=\int_{\alpha_{s}\left(\mu_{0}\right)}^{\alpha_{s}(\mu)}
d \alpha \frac{\Gamma_{\mathrm{cusp}}(\alpha)}{\beta(\alpha)}\, .
\end{eqnarray}
\item Performing the Mellin transformation to the evolution equation in the position space \cite{Kawamura:2010tj,Braun:2019zhp}:
\begin{align}
\tilde\varphi^+(j,\mu) =&~ \frac{1}{2\pi i} \int^{-i\infty}_{-i0} \frac{dt}{t} \,
(it\tilde \mu )^{-j} \,
\tilde\Phi^+(t,\mu) \,.
\end{align}

In the Mellin space, the evolution equation takes a simple form:
\begin{align}\label{mellinequation}
\Big[\frac{d}{d\ln\mu}
+\hat V(j,\alpha_s)\Big]\,\tilde\varphi^+(j,\mu)
=0\,,
\end{align}
with
\begin{align}
\hat V(j,\alpha_s)
=j+\gamma_+-\gamma_F+
\Gamma_{\rm cusp}\,\Big[\psi(j+2)-\psi(2)+\vartheta(j)\Big]\,,
\end{align}
where $\vartheta(j)=0$ at the one-loop level. The solution in the Mellin space can be obtained directly:
\begin{align}\label{mellinsolution}
\tilde\varphi^+(j(\mu),\alpha_s(\mu),\mu)=\tilde\varphi^+(j(\mu_0),\alpha_s(\mu_0),\mu_0)\exp
\left\{-\int_{\mu_0}^\mu{ds\over s}\hat V[j(s),\alpha_s(s)]\right\}\, .
\end{align}
\item  It was found that, if the $B$-meson LCDA is transformed into the so-called ``dual" space, the evolution kernel is diagonalized \cite{Bell:2013tfa}. The LCDA in the dual space can be obtained by
\begin{eqnarray}\label{dual}
\rho^{+}\left(\omega^{\prime}, \mu\right)=\int_{0}^{\infty} \frac{d \omega}{\omega}
\sqrt{\frac{\omega}{\omega^{\prime}}} J_{1}\left(2 \sqrt{\frac{\omega}{\omega^{\prime}}}\right)
\phi^{+}(\omega, \mu),
\end{eqnarray}
which satisfies an ordinary differential equation:
\begin{eqnarray}
\mu \, {d \over d \mu} \,\rho^+ (\omega',\mu) = -\Big[\Gamma_{\rm
cusp}\ln{\mu\over \hat{\omega'}}+\gamma \Big]\rho^+ (\omega',\mu)\, .
\end{eqnarray}
It is then simple to write the solution:
\begin{eqnarray}
\rho^{+}\left(\omega^{\prime}, \mu\right)
=e^{V}\left(\frac{\mu_{0}}{\hat{\omega}^{\prime}}\right)^{-g}
 \rho^{+}\left(\omega^{\prime}, \mu_{0}\right)
 =e^{\bar{V}}\left(\frac{\mu \mu_{0}}{\left(\hat{\omega}^{\prime}\right)^{2}}\right)^{-g / 2}
 \rho^{+}\left(\omega^{\prime}, \mu_{0}\right) \, ,
\end{eqnarray}
with
\begin{align}
&\hat{\omega}^{\prime}
=e^{-2 \gamma_{E}} \omega^{\prime},
&&\bar{V}\left(\mu, \mu_{0}\right)
=\frac{1}{2}\left(V\left(\mu, \mu_{0}\right)-V\left(\mu_{0}, \mu\right)\right)\, .\nonumber
\end{align}
\end{itemize}
The method mentioned above is equivalent, and the LCDAs $\phi^+(\omega), ~\varphi^+(\theta), ~\tilde \Phi^+(t), ~\tilde \varphi^+(j),~ \rho^+(\omega')$  are different expressions of an identical objective.
Because the momentum space and the position space are related through a standard Fourier transformation, we are able to derive
\begin{eqnarray}\label{MellMom}
\tilde\varphi_+(j) = \frac{\Gamma(-j)}{2\pi i} \, \int^{\infty}_{0} d\omega\,
\Big(\frac{\omega}{\tilde\mu }\Big)^j \, \phi_+(\omega) \,
\end{eqnarray}
and
\begin{eqnarray}
\tilde\varphi^+(-i\theta) = {\Gamma(i\theta)\over 2\pi i}\,
{\mu^{1-i\theta}\over\tilde\mu^{-i\theta}} \, \varphi^+(\theta+i).
\end{eqnarray}
$\rho^+(\omega')$ is related to $\varphi^+(\theta)$ by definition:
\begin{eqnarray}
\varphi^+(\theta,\mu)={\Gamma(1-i\theta)\over \Gamma(1+i\theta)}\int_0^\infty {d\omega'\over \omega'}\rho^+(\omega',\mu)\left({\mu\over \omega'}\right)^{i\theta}\, .\end{eqnarray}
Then, we have
\begin{eqnarray}
\tilde\varphi^+(j,\mu)={\tilde \mu\over 2\pi i} \, {\Gamma(2+j) }\int_0^\infty {d\omega'\over \omega'}\rho^+(\omega',\mu)\left({\tilde\mu\over \omega'}\right)^{-1-j}\, .\end{eqnarray}
At the one-loop level, the most convenient method  is to work in the dual space since the Bessel function is the eigenfunction of the Lange-Neubert kernel, which is confirmed in \cite{Braun:2014owa,Braun:2018fiz}. The Lange-Neubert kernel can be expressed as a logarithm of the generator of special conformal transformations along the light cone. When the eigenfunction of the generator is transformed to the momentum space, it is simply the Bessel function in Eq. (\ref{dual}).

The two-loop-level anomalous dimension of the $B$-meson LCDA was first calculated in the coordinate space in \cite{Braun:2019wyx}; it is more simply expressed in the dual space:
\begin{eqnarray}
\left[\mu{\partial\over \partial \mu} +\beta(a){\partial\over \partial{a}}+\Gamma_{\rm cusp}(a)\ln(\tilde \mu e^{\gamma_E}s)+\gamma(a)\right]\eta_+(s,\mu)=4\,C_F\,a^2
\int_0^1{du\over u} \,\bar u \,h(u)\,\eta_+(\bar us,\mu)\, ,
\end{eqnarray}
where $s\eta_+(s)=\rho^+(1/s)$ and $a=\alpha_s/(4\pi)$. This equation is also transformed into the momentum space in \cite{Liu:2020ydl}, resulting in the two-loop-level Lange-Neubert equation. The advantage of solving the evolution equation at the two-loop level in the dual space does not hold as the two-loop evolution kernel is not diagonal in this space. On the contrary, the elegant form of the evolution equation (Eq. (\ref{mellinequation})) in the Mellin space  is maintained.
Thus, Eq.(\ref{mellinsolution}) is still the solution to the evolution equation up to the two-loop level with \cite{Braun:2019zhp}
\begin{align}
  \vartheta(j) &= a  \vartheta^{(1)}(j) = a \biggl\{(\beta_0-3C_F)\Big(\psi^\prime(j+2)-\psi^\prime(2)\Big)
+2C_F\biggl(\frac1{(j+1)^3}
 \notag\\
&\quad
+\psi^\prime(j+2)(\psi(j+2)-\psi(1))+\psi^\prime(j+1)(\psi(j+1)-\psi(1))-\frac{\pi^2}6\biggr)
  \biggr\},
  \notag\\
  \gamma_+(a) &=
   -a C_F +  a^2 C_F
\biggl\{
4 C_F \left[\frac{21}{8} + \frac{\pi^2}{3} - 6\zeta_3\right]
+ C_A \left[\frac{83}{9} -\frac{2\pi^2}{3} - 6\zeta_3\right]
+ \beta_0\left[\frac{35}{18} -\frac{\pi^2}{6}\right]
\biggr\}\,,
\nonumber \\
\gamma_F(a) &= -3 a C_F + a^2 C_F\bigg\{C_F\Big[\frac{5}{2}-\frac{8\pi^2}{3}\Big]
+ C_A \Big[1+ \frac{2\pi^2}{3}\Big] -\frac{5}{2} \beta_0\bigg\}.
\end{align}

In a recent paper \cite{Galda:2020epp}, an alternative approach to solving the evolution equation at the two-loop level was proposed. The essential idea of this approach is to perform a Laplace transformation on the $B$-meson LCDA,
\begin{align}
\tilde\phi^+(\eta,\mu) = \int^\infty_0 \frac{d\omega}{\omega}\,
\Big(\frac{\omega}{\bar \omega}\Big)^{-\eta} \, \phi^+(\omega,\mu) \,,
\end{align}
where $\bar\omega$ is a fixed reference scale, which can be used to eliminate the logarithmic moment $\sigma_1$ in the factorization formula of $B \to \gamma\nu\ell$.
Then, one could derive
\begin{equation}\label{Lapeq}
\begin{aligned}
   & \left( \frac{d}{d\ln\mu}
    + \Gamma_{\rm cusp}(\alpha_s)\,\frac{\partial}{\partial\eta} \right) \tilde\phi_+(\eta,\mu) = \bigg[ \Gamma_{\rm cusp}(\alpha_s) \left( \!\ln\frac{\bar\omega}{\mu} + {\cal F}(\eta)\! \right)
    - \gamma(\alpha_s) + {\cal G}(\eta,\alpha_s) \bigg]\,\tilde\phi_+(\eta,\mu) \,,
\end{aligned}
\end{equation}
with the definition
\begin{align}\label{calGdef}
   \F(\eta)
   &= \int_0^\infty\!dx\,\Gamma(1,x)\,x^\eta = - \big[ H(\eta) + H(-\eta) \big] \,, \notag \\
   \G(\eta;\alpha_s)
   &= \int_0^\infty\!dx\,\hat\gamma_+(1,x;\alpha_s)\,x^\eta \,,
\end{align}
where $\hat\gamma_+(1,x;\alpha_s)$ starts from the two-loop level, and the specific expression can be seen in \cite{Galda:2020epp}. After the Laplace transformation, the solution to the evolution equation reads
\begin{align}\label{Lapsolu}
   \tilde\phi_+(\eta,\mu)
   &= N(\mu_s,\mu)\,
    \frac{\Gamma\big(1+\eta+a_\Gamma(\mu_s,\mu)\big)\,\Gamma(1-\eta)}%
         {\Gamma\big(1-\eta-a_\Gamma(\mu_s,\mu)\big)\,\Gamma(1+\eta)} \notag \exp\Bigg[\,\int\limits_{\alpha_s(\mu_s)}^{\alpha_s(\mu)}\!\frac{d\alpha}{\beta(\alpha)}\,
    {\cal G}\big(\eta+a_\Gamma(\mu_\alpha,\mu),\alpha\big) \Bigg] \notag \\[2mm]
   &\times e^{2\gamma_E a_\Gamma(\mu_s,\mu)}\,\tilde\phi_+\big(\eta+a_\Gamma(\mu_s,\mu),\mu_s\big) \,.
\end{align}
The normalization $N(\mu_s,\mu)$ depends on the factorization scale through
\begin{equation}
   N(\mu_s,\mu)
   = \left( \frac{\bar\omega}{\mu_s} \right)^{-a_\Gamma(\mu_s,\mu)}
    e^{\,S(\mu_s,\mu) + a_\gamma(\mu_s,\mu)} \,,
\end{equation}
where the quantities $a_\gamma, a_\Gamma$,  and $S(\mu_s,\mu)$ are given as
\begin{eqnarray}\label{Sdef}
S(\mu_s,\mu)
   &=& - \int\limits_{\alpha_s(\mu_s)}^{\alpha_s(\mu)}\!d\alpha\,
    \frac{\Gamma_{\rm cusp}(\alpha)}{\beta(\alpha)}
    \int\limits_{\alpha_s(\mu_s)}^\alpha\!\frac{d\alpha'}{\beta(\alpha')} \, ,\nonumber\\
    a_\Gamma(\mu_0,\mu)
   &=& - \int\limits_{\alpha_s(\mu_0)}^{\alpha_s(\mu)}\!
    d\alpha\,\frac{\Gamma_{\rm cusp}(\alpha)}{\beta(\alpha)}
     ,\,\,\,
a_\gamma(\mu_0,\mu)
   = - \int\limits_{\alpha_s(\mu_0)}^{\alpha_s(\mu)}\!
    d\alpha\,\frac{\gamma(\alpha)}{\beta(\alpha)}\,.
  \end{eqnarray}
We note that the LCDA $ \varphi^+(j)$ is related to $\tilde\phi^+(\eta)$ through \cite{Shen:2020hfq}
\begin{align}
\tilde\varphi^+(j,\mu)
=&~\frac{\Gamma(-j)}{2\pi i} \,\tilde \mu\,
\Big(\frac{\bar \omega}{\tilde \mu}\Big)^{j+1}\, \,
\tilde\phi^+(-j-1,\mu) \,.
\end{align}

\section{ $B \to \gamma $ form factors }
At leading power in $\Lambda/m_b$, the QCD factorization formula for the $B \to \gamma$
form factors can be written as
\begin{eqnarray}
F_{V,\, \rm LP}(E_\gamma) = F_{A, \, \rm LP} (E_\gamma)= {Q_u \, m_B \over 2E_\gamma} \, \tilde{f}_B(\mu) \,
C_{\perp}(E_\gamma, \mu) \, \int_0^{\infty} \, d \omega \, {\phi_B^{+}(\omega, \mu) \over \omega} \,
J_{\perp}(E_\gamma,\omega,  \mu) \, .
\label{leading-power factorization formula}
\end{eqnarray}
At the one-loop level, the hard function and jet function are given as
\cite{Bauer:2000yr,Lunghi:2002ju,Bosch:2003fc}
\begin{eqnarray}
C_{\perp}(E_\gamma,\mu)&=& 1- \frac{\alpha_s \, C_F}{4 \, \pi}
\bigg [ 2 \, \ln^2 {\mu \over 2E_\gamma} + 5 \, \ln {\mu \over m_b}
-2 \, {\rm Li}_2 \left ( 1-{1 \over r} \right )  - \ln^2  r \, , \nonumber \\
&&  + \,  {3 r -2 \over 1 -r}  \, \ln r + {\pi^2 \over 12} + 6  \bigg ]+ {\cal O}(\alpha_s^2) \nonumber \\
J_{\perp}(E_\gamma,\omega,\mu)&=&  1 + {\alpha_s \, C_F \over 4 \, \pi} \,
\left [ \ln^2 { \mu^2 \over 2E_\gamma \,  \omega }  - {\pi^2 \over 6} - 1 \right ]
+ {\cal O}(\alpha_s^2) \,,
\end{eqnarray}
with $r=2E_\gamma /m_b$. The results of the two-loop level hard function and jet function can be found in  \cite{Beneke:2011nf,Liu:2020ydl}. As the hard function and jet function contain large logarithmic terms, it is important to perform resummation to improve the convergence of the perturbative series.
The first complete NLL resummation is given in \cite{Galda:2020epp}:
\begin{equation}\label{master}
\begin{aligned}
   &F_{A/V,\rm LP}(E_\gamma)
   \\
  =& {Q_u \, m_B \over 2E_\gamma\,\, }\,\exp\Big[ S(\mu_h,\mu_j) + S(\mu_s,\mu_j) - a_{\gamma_H}(\mu_h,\mu_j)
    + a_\gamma(\mu_s,\mu_j) + 2\gamma_E\hspace{0.3mm}a_\Gamma(\mu_s,\mu_j) \Big] \\[1mm]
   &\times
   \tilde f_B(\mu_h) \,C_{\perp}(E_\gamma,\mu_h) \left( \frac{2E_\gamma}{\mu_h} \right)^{-a_\Gamma(\mu_h,\mu_j)}
    \J(\partial_\eta,\mu_j)\,\bigg( \frac{2E_\gamma\hspace{0.2mm}\bar\omega}{\mu_j^2} \bigg)^\eta\,\,
    \frac{\Gamma\big(1-\eta+a_\Gamma(\mu_s,\mu_j)\big)\,\Gamma(1+\eta)}%
         {\Gamma\big(1+\eta-a_\Gamma(\mu_s,\mu_j)\big)\,\Gamma(1-\eta)} \\[-1mm]
   &\times
    \exp\Bigg[\,\int\limits_{\alpha_s(\mu_s)}^{\alpha_s(\mu_j)}\!
    \frac{d\alpha}{\beta(\alpha)}\,\G\big(-\eta+a_\Gamma(\mu_\alpha,\mu_j),\alpha\big) \Bigg]
    \left( \frac{\bar\omega}{\mu_s} \right)^{-a_\Gamma(\mu_s,\mu_j)}
    \tilde\phi_+\big(\!-\!\eta+a_\Gamma(\mu_s,\mu_j),\mu_s\big)\, \bigg|_{\eta=0} \,,
\end{aligned}
\end{equation}
where the jet function $\J(L_p,\mu_j)\equiv J_\perp(-p^2,\mu_j)=J_\perp(n\cdot p\omega,\mu_j)$ with $L_p=\ln(p^2/\mu_j^2)$.
We now derive the scale independent factorization formula at the NLL level in the Mellin space. The evolution of the $B$-meson LCDA is known in the Mellin space. We thus need to perform the Mellin transformation to the jet function, although it is not well-defined. Alternatively, we follow the method in \cite{Galda:2020epp} to replace the first argument of $\J(L_p,\mu_j)$ by a derivative operator, i.e.,
\begin{eqnarray}
 \int_0^{\infty} \, {d \omega \over \omega} \,\phi^{+}(\omega, \mu) \,
J_{\perp}(E_\gamma,\omega,  \mu)
&=&{\J}_{\perp}(\partial_j,\mu)\left.\left( 2E_\gamma  \over \mu\right)^j\int_0^{\infty} \, {d \omega \over \omega} \, {\phi^{+}(\omega, \mu)} \,
\left( \omega \over \mu\right)^j\right |_{j=0}\, .
\end{eqnarray}
Taking advantage of  Eq. (\ref{MellMom}), we  have
\begin{eqnarray}
 \int_0^{\infty} \, {d \omega \over \omega} \,\phi^{+}(\omega, \mu) \,
J_{\perp}(E_\gamma,\omega,  \mu)
&=&2\pi \,i\,{\J}_{\perp}(\partial_j,\mu)\left.
\frac{1}{\tilde \mu}\left( 2E_\gamma e^{2\gamma_E}  \over \mu\right)^j{1\over \Gamma(1-j)} \tilde\varphi^+(j-1,\mu)\right |_{j=0}\, .
\end{eqnarray}
Employing the evolution function of the hard function and $B$-meson LCDA, we  obtain
\begin{eqnarray}\label{master2}
   F_{A/V,\rm LP}
  &=&{Q_u \, m_B \over 2E_\gamma \,\, } \,
\left [  U_2(E_\gamma, \mu'_{h}, \mu) \,\tilde{f}_B(\mu'_{h}) \right ] \,
\left [ U_1(E_\gamma, \mu_{h}, \mu) \, C_{\perp}(E_\gamma, \mu_{h})   \right ]\times 2\pi \,i\,{\J}_{\perp}(\partial_j,\mu_j)\,\frac{1}{\tilde \mu}
    \nonumber \\
   &&
   \left.\left( 2E_\gamma e^{2\gamma_E} \over \mu\right)^j{1\over \Gamma(1-j)} \tilde\varphi^+(j(\mu_s)-1,\alpha_s(\mu_s),\mu_s)\exp
\left\{-\int_{\mu_s}^\mu{ds\over s}V[j(s),\alpha_s(s)]\right\}\right |_{j=0}\, ,
\end{eqnarray}
where $U_1(E_\gamma, \mu_{h}, \mu)$ and $U_2(E_\gamma, \mu'_{h}, \mu) $ are the evolution factor of the hard function and $B$-meson decay constant in the HQET, respectively; the specific expression can be found in the appendices of \cite{Beneke:2011nf,Wang:2016qii}. The parameter $j$ depends on the factorization scale through
\begin{eqnarray}
\mu{d\over d\mu}j(\mu)=-\Gamma_{\rm cusp}{(\alpha_s)}.\end{eqnarray}
The resummed factorization formula in the Mellin space seems more compact than Eq. (\ref{master}).

Now, we turn to the power suppressed contributions. At leading  power, $F_V$ and $F_A$ are equal due to the left-handedness of the weak interaction current and the helicity-conservation of the quark-gluon interaction in the high-energy limit, although this relation might be broken by the power corrections.  In \cite{weiyb}, the power suppressed contribution is separated into the symmetry-preserving part $\xi$ and symmetry-breaking part $\Delta \xi$, i.e.,
\begin{align}
   F^{\rm NLP}_{V}(E_\gamma)&=\xi(E_\gamma)+\Delta \xi(E_\gamma) \, ,\nonumber \\
   F^{\rm NLP}_{A}(E_\gamma)&=\xi(E_\gamma)-\Delta \xi(E_\gamma)\, .
   \label{xi}
\end{align}
When the power suppressed contribution is from the region where $x^2\sim 1/\Lambda^2$, in which $x$ denotes the separation between the quark-photon vertex and the weak current, it is called the soft contribution. Soft contributions  with higher twist $B$-meson LCDA are considered in \cite{weiyb}.  Because they are highly suppressed and numerically small, we neglect them in this study. Then, the symmetry breaking part only contains the local contribution and can be written as
\begin{eqnarray}
\Delta \xi(E_\gamma)={e_uf_Bm_B\over 4E_\gamma^2}+{e_bf_Bm_B\over 2E_\gamma m_b}\, .
\end{eqnarray}
The symmetry preserving part can be divided into three parts, i.e., $\xi(E_\gamma)=\xi_{1\over E_\gamma}(E_\gamma)+\xi_{1\over m_b}(E_\gamma)+\xi_{soft}(E_\gamma)$, and the explicit expressions for the first two parts are
\begin{eqnarray}
\xi_{1\over E_\gamma}(E_\gamma)&=&{e_uf_Bm_B\over 4E_\gamma^2}\left[-1+2\int_0^\infty d\omega\ln\omega\phi^-_{t3}(\omega)-2\int_0^\infty{d\omega_2\over \omega_2}\phi_4(0,\omega_2)\right]\, ,\nonumber \\
\xi_{1\over m_b}(E_\gamma)&=&{e_uf_Bm_B\over 4E_\gamma m_b}\left[{\bar \Lambda \over \lambda_B}-2+2\int_0^\infty {d\omega_1\over \omega_1}\int_0^\infty{d\omega_2\over \omega_1+\omega_2}\phi_3(\omega_1,\omega_2)\right]\, ,
\end{eqnarray}
where $\phi_3(\omega_1, \omega_2)$ and $\phi_4(\omega_1, \omega_2)$ are the three particle twist-3 and twist-4 $B$-meson LCDAs, respectively. $\phi^-_{t3}(\omega)$ is the  ``genuine" twist-three contribution to the LCDA $\phi^-(\omega)$ \cite{weiyb}. The soft contribution with QCD corrections is obtained as
\begin{eqnarray}
\xi_{soft}(E_\gamma)&=&{e_uf_Bm_B\over 2E_\gamma}C_\perp(E_\gamma,\mu_{h})K^{-1}(\mu'_{h})U(E_\gamma,\mu_{h},\mu'_{h},\mu)\nonumber \\ &\times&\int_0^{s_0\over 2E_\gamma}d\omega'\left[{2E_\gamma\over m_\rho^2}e^{-{2E_\gamma\omega'-m_\rho^2\over M^2}}-{1\over \omega'}\right]\rho^+_{\rm eff}(\omega',\mu)\, ,
\end{eqnarray}
where $K(\mu)$ is the factor relating the QCD decay constant of the $B$-meson to the HQET one,  and $M^2$ and $s_0$ are theBorel mass and threshold parameter, respectively. The evolution kernel $U(E_\gamma,\mu_{h},\mu'_{h},\mu)=U_1(E_\gamma,\mu_{h},\mu)/U_2(E_\gamma,\mu'_{h},\mu)$. The effective LCDA $\rho^+_{\rm eff}(\omega',\mu)$ takes the form
\begin{eqnarray}
\rho^+_{\mathrm{eff}}(\omega',\mu)&=&\phi^+(\omega',\mu)+{\alpha_sC_F\over 4\pi}\Bigg\{\left(\ln^2{\mu^2\over 2E_\gamma \omega'}+{\pi^2\over 6}-1\right)\phi^+(\omega',\mu)\nonumber \\
&+&\left(2\ln{\mu^2\over 2E_\gamma \omega'}+3\right)\omega'\int_{\omega'}^\infty d\omega \ln{\omega-\omega'\over \omega'}{d\over d\omega}{\phi^+(\omega,\mu)\over \omega}
\nonumber \\
&-&2\ln{\mu^2\over 2E_\gamma \omega'}\int_0^{\omega'} d\omega \ln{\omega'-\omega\over \omega'}{d\over d\omega}{\phi^+(\omega,\mu)}
\nonumber \\
&+&\int_0^{\omega'} d\omega \ln^2{\omega'-\omega\over \omega'}{d\over d\omega}{\left[{\omega'\over \omega}\phi^+(\omega,\mu)+\phi^+(\omega,\mu)\right]}
\Bigg\}\, .
\end{eqnarray}
To obtain this function, one must generalize the photon momentum from $p^2=0$ to $-p^2\neq 0$, calculate the generalized hard-collinear function in this Euclidean region, perform the dispersion treatment to the convolution of the hard-collinear function with the $B$-meson LCDA, and finally take the limit $p^2\to 0$. The soft contribution actually includes the hadronic effect of the photon; it must overlap with the contribution of the photon LCDA, which will be investigated in future work.

\section{Phenomenological analysis }
\begin{figure}
\centering
\includegraphics[clip,width=0.6\textwidth]{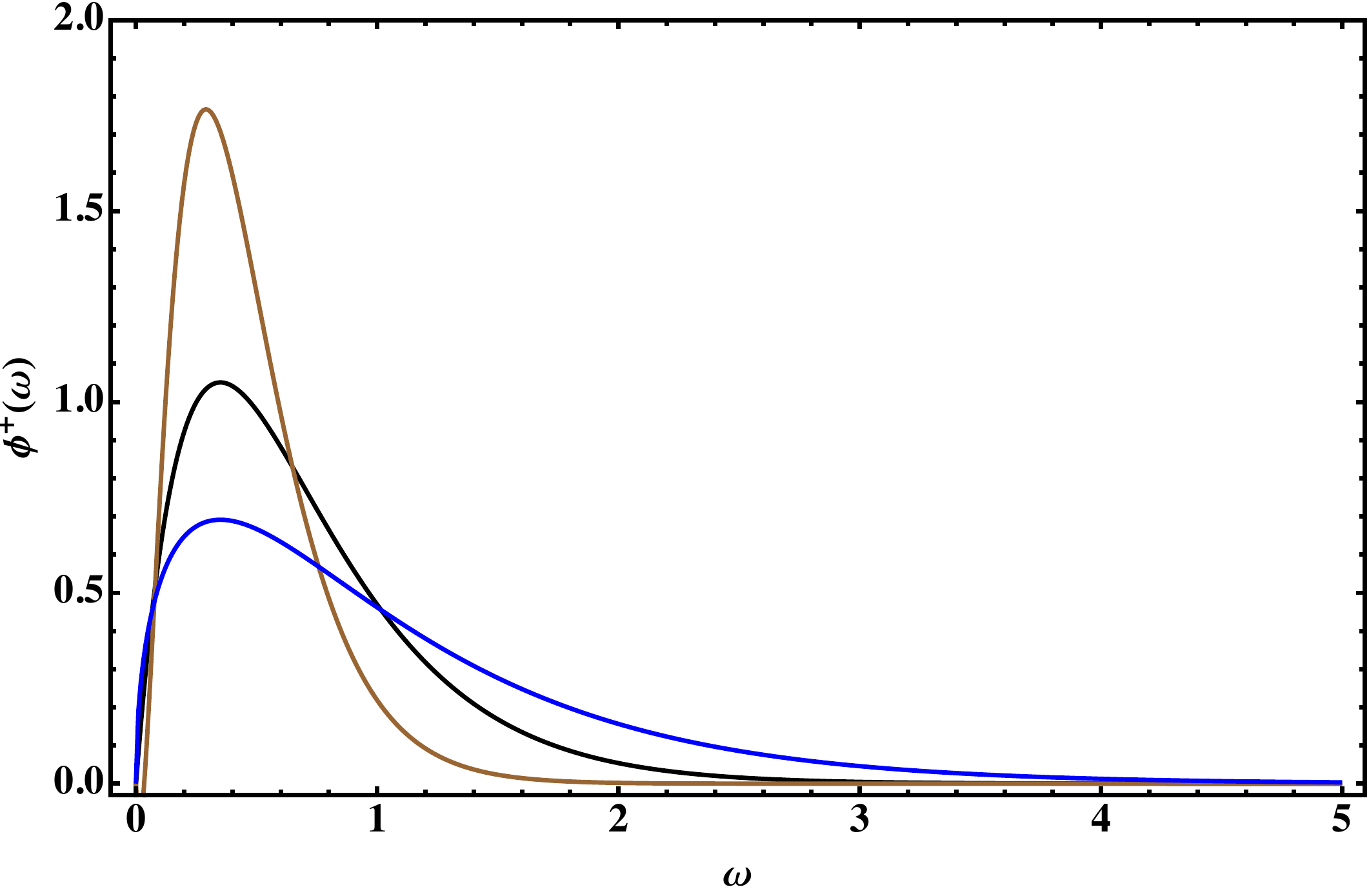}
\caption{The leading-twist $B$-meson LCDA with the three-parameter model.
Setting $\lambda_B=0.35$ GeV, the black curve represents the exponential model, i.e., $\beta=\alpha$. The brown and blue curves represent $\alpha=2.0,\beta=1.5$ and $\alpha=1.5,\beta=2.0$, respectively. } \label{fig:lcda}
\end{figure}

The fundamental nonperturbative inputs entering the factorization formula of $B \to \gamma \nu \ell$ decay include the two-particle and
three-particle $B$-meson distribution amplitudes up to the
twist-four accuracy. The decay constant of the $B$-meson and the parameters appear in the dispersion approach. In the numerical analysis, we employ the following three-parameter model for the leading twist $B$-meson LCDA \cite{weiyb}
\begin{eqnarray}
\phi^+(\omega)={\Gamma(\beta)\over \Gamma(\alpha)}\,{\omega\over \omega^2_0}\,e^{-{\omega\over \omega_0}}\,U\left(\beta-\alpha,3-\alpha,{\omega\over \omega_0}\right),\end{eqnarray}
where $U(\alpha,\gamma,x)$ is the confluent hypergeometric function of the second kind. In dual space, this model has a simpler expression:
\begin{eqnarray}
\rho^+(\omega')={1\over \omega'}\,{}_1F_1\left(\alpha,\beta,-{\omega_0\over \omega'}\right),\end{eqnarray}
where ${}_1F_1\left(\alpha,\beta,z\right)$ is the  confluent hypergeometric function of the first kind.
In the leading power factorization formula, only the first inverse moment and the logarithmic moments enter the factorization formula; they are defined by
\begin{eqnarray}\label{moment}
{1\over\lambda_B(\mu)}&=&\int_0^\infty{d\omega\over \omega}\phi^+(\omega)\, ,\nonumber \\
\sigma_n(\mu)&=&\lambda_B(\mu)\int_0^\infty{d\omega\over \omega}\ln^n{\mu_0\over \omega}\phi^+(\omega) \, .\end{eqnarray}
For the three-parameter model,  the first inverse moment and the first two logarithmic moments are obtained  as
\begin{eqnarray}
\lambda_B&=&{\alpha-1\over \beta-1}\omega_0\, ,\nonumber \\
\sigma_1&=&\psi(\beta-1)-\psi(\alpha-1)+\ln{\mu_0\over \omega_0}+\gamma_E\, ,\nonumber \\
\sigma_2&=&\sigma_1^2+\psi'(\beta-1)-\psi'(\alpha-1)+{\pi^2\over 6}\, .\end{eqnarray}
 If the parameter $\alpha=\beta$, the three-parameter model is simply the familiar exponential model
\cite{Grozin:1996pq}
 \begin{eqnarray}
\phi^+(\omega)={\omega\over \omega^2_0}\,e^{-{\omega\over \omega_0}},\end{eqnarray}
which is set as our default model. To estimate the error from the models, we let $\alpha-\beta$ vary in the region $-0.5<\alpha-\beta<0.5$. We then employ two models with
 $\alpha=2.0, \beta=1.5$ and $\alpha=1.5, \beta=2.0$. For the default model, $\omega_0=\lambda_B$,
 whose determination
has been discussed extensively in the context of exclusive $B$-meson decays
(see \cite{Wang:2015vgv,Wang:2017jow,Lu:2018cfc,Gao:2019lta} for further discussion). Here, we employ
$\lambda_B(1 \, {\rm GeV})=0.35\pm0.05 \, {\rm GeV}$, which is consistent with the calculations of the semileptonic
$B \to \pi$ form factors with $B$-meson LCDAs in the framework of light-cone sum rules \cite{Wang:2015vgv}. The leading twist $B$-meson LCDA with the three-parameter model is plotted in Fig. \ref{fig:lcda}. In the factorization formula (\ref{Lapsolu}), a new parameter $\bar \omega$ is introduced to eliminate $\sigma_1$; it can be determined once the parameters $\alpha, \beta$, and $\lambda_B$ are given.  In addition,  by utilizing the three-parameter $B$-meson LCDA model, the logarithmic moments $\sigma_{2,3,4}$ are also determined. All the parameters are listed in Table \ref{tab of parameters for B meson DAs}.
\begin{table}[h!]
\begin{center}
\setlength\tabcolsep{8pt}
\def\arraystretch{1.8}
\begin{tabular}{|c|c|c|c|c|c|c|c|}
  \hline
  \hline
  Model&$\alpha$&$\beta$&$\omega_0$ [GeV]&$\sigma_2$&$\sigma_3$&$\sigma_4$&$\lambda_B$ [GeV]
 \\
 \hline
 Default& arbitrary &$\beta=\alpha$&$0.350$&$1.64$&$2.4$&$14.6$&$0.35$
 \\
 \hline
 Model 1&2.0&1.5&$0.175$&$-1.64$&$-12.2$&-76.3&$0.35$
 \\
 \hline
 Model 2&1.5&2.0&$0.700$&$4.93$&$16.8$&$170.0$&$0.35$
 \\
\hline
\hline
\end{tabular}
\end{center}
\caption{Numerical values of the nonperturbative parameters entering the leading twist LCDA of the $B$ meson. Here, the energy scale is $\mu_s=1~{\rm GeV}$, and $\lambda_B$ is fixed at 0.35 GeV.}
\label{tab of parameters for B meson DAs}
\end{table}

The higher-twist LCDAs must incorporate
the correct low-momentum behaviour and satisfy the equations of motion. All the suggested  models can be obtained as particular cases of a more general ansatz:
\begin{eqnarray}
\phi^+(\omega)&=&\omega \, f(\omega)\, ,\nonumber \\
\phi_3(\omega_1,\omega_2)&=&-{1\over 2}\, \mathcal{N}\,(\lambda_E^2-\lambda_H^2)\,\omega_1\,\omega_2^2\,f'(\omega_1+\omega_2)\, ,\nonumber \\
\phi_4(\omega_1,\omega_2)&=&{1\over 2}\,\mathcal{N}\,(\lambda_E^2+\lambda_H^2)\,\omega^2_2\,f(\omega_1+\omega_2)\, .\end{eqnarray}
Here, the function $f(\omega)$ obeys the following normalization condition:
\begin{align}
&\int_0^\infty \omega \, f(\omega) \,d\omega=1\, ,\,\,\,\,
&&{1\over \mathcal{N}}={1\over 2}\int_0^\infty \omega^3 \,f(\omega)\,d\omega=\bar \Lambda^2+{1\over 6}(2\lambda_E^2-\lambda_H^2)\, .
\end{align}
The following results can then be derived:
\begin{eqnarray}
\int_0^\infty{d\omega\over \omega}\,\ln\omega\,\phi^{t3}_-(\omega)&=&{1\over 6}\,\mathcal{N}\,(\lambda_E^2-\lambda_H^2)\, ,\nonumber \\
\int_0^\infty {d\omega_1\over \omega_1}\int_0^\infty{d\omega_2\over \omega_1+\omega_2}\,\phi_3(\omega_1,\omega_2)&=&{1\over 3}\,\mathcal{N}\,(\lambda_E^2-\lambda_H^2)\, ,\nonumber \\
\int_0^\infty {d\omega_2\over \omega_2}\,\phi_4(0,\omega_2)&=&{1\over 2}\,\mathcal{N}\,(\lambda_E^2+\lambda_H^2)\, .\end{eqnarray}
Taking advantage of the above results, the NLP contribution with $1/E_\gamma$ and $1/m_b$ corrections can be obtained as
\begin{eqnarray}
\xi_{1\over E_\gamma}(E_\gamma)&=&-{e_uf_Bm_B\over 2E_\gamma^2}\left[{1\over 2}+{2(\lambda_E^2+2\lambda_H^2)\over 6{\bar\Lambda}^2+2\lambda_E^2+\lambda_H^2}\right]\, ,\nonumber \\
\xi_{1\over m_b}(E_\gamma)&=&{e_uf_Bm_B\over 4E_\gamma m_b}\left[{\bar \Lambda \over \lambda_B}-2+{4(\lambda_E^2-\lambda_H^2)\over 6{\bar\Lambda}^2+2\lambda_E^2+\lambda_H^2}\right]\, .
\end{eqnarray}
To highlight the influence of the power suppressed contributions, we present the numerical results of the form factors with different contributions:
\begin{eqnarray}
F_A&=&{1\over \lambda_B}\left(0.102+0.0067\sigma_2+5.4\times 10^{-5}\sigma_3+7.6\times 10^{-5}\sigma_4\right)\big|_{\rm LP}+\left(-0.135+{0.00768\over \lambda_B}\right)\bigg|_{\rm NLP}\, ,\nonumber \\
F_V&=&{1\over \lambda_B}\left(0.102+0.0067\sigma_2+5.4\times 10^{-5}\sigma_3+7.6\times 10^{-5}\sigma_4\right)\big|_{\rm LP}+\left(-0.101+{0.00768\over \lambda_B}\right)\bigg|_{\rm NLP}\ ,
\end{eqnarray}
where the photon energy $E_\gamma$ is fixed at $2.2$ GeV, and $\lambda_B$ and $\sigma_n$ are set to be free parameters. The leading power result in the first parentheses is borrowed from \cite{Galda:2020epp}, and the power suppressed contributions include the symmetry breaking local term, the symmetry preserving $1/E_\gamma, 1/m_b$ term, and the soft contribution.
It is obvious that power corrections are sizeable and more important than the $\sigma_n$ terms. Therefore, the power suppressed contribution must play an important role in the determination of $\lambda_B$. We leave a more detailed study of the subleading power corrections for a future work.
\begin{figure}
\centering
\includegraphics[clip,width=0.6\textwidth]{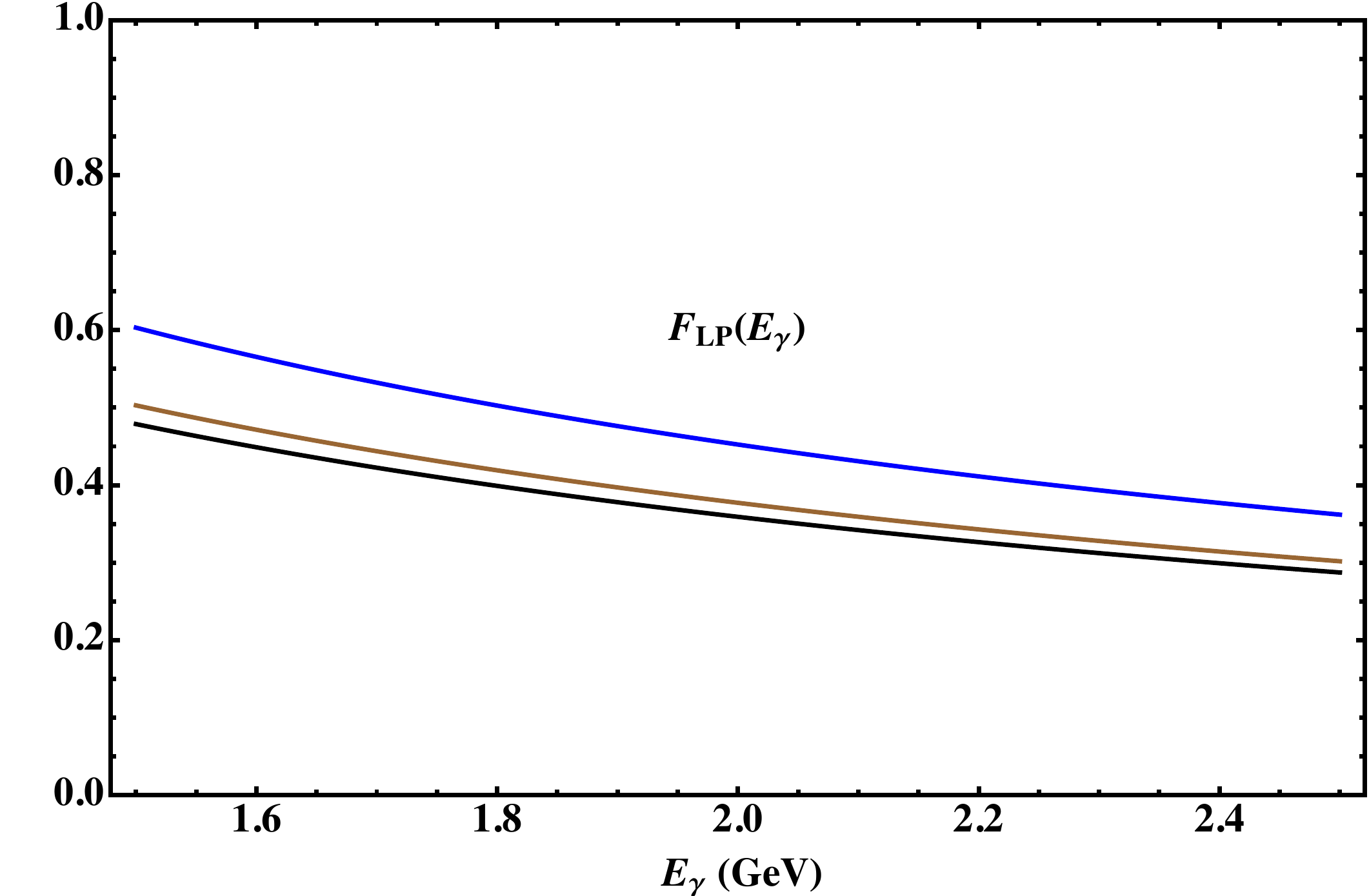}
\caption{The leading power contribution to the form factors. The blue, brown and black curves stand for the LL, NLL and NNLO results, respectively. } \label{fig:fLL}
\end{figure}

To test the effect of the large logarithm resummation, we plot the $E_\gamma$ dependence of the  leading power form factor $F_{A/V,\rm LP}$ in Fig. \ref{fig:fLL}. For the leading logarithmic resummation, we employ the tree-level hard function and jet function as well as the one-loop level anomalous dimension and two-loop cusp anomalous dimension. For the NLL resummation case, we follow the convention in \cite{Galda:2020epp}. In \cite{Galda:2020epp}, the contributions from the NNLO hard function and jet function are also considered. Strictly speaking, to resum the logarithmic terms in the NNLO jet function, we need the three-loop anomalous dimension of the $B$-meson LCDA, which has not yet been obtained. While it is not phenomenologically important, as the hard-collinear scale $\mu_j$ is actually close to the soft scale $\mu_s$.
From Fig. \ref{fig:fLL}, we can see that the NLL resummation effect significantly decreases the LL result, and that the NNLO result is approximately 5$\%$ smaller than the NLL result.

\begin{figure}[h!]
\centering
\includegraphics[clip,width=0.48\textwidth]{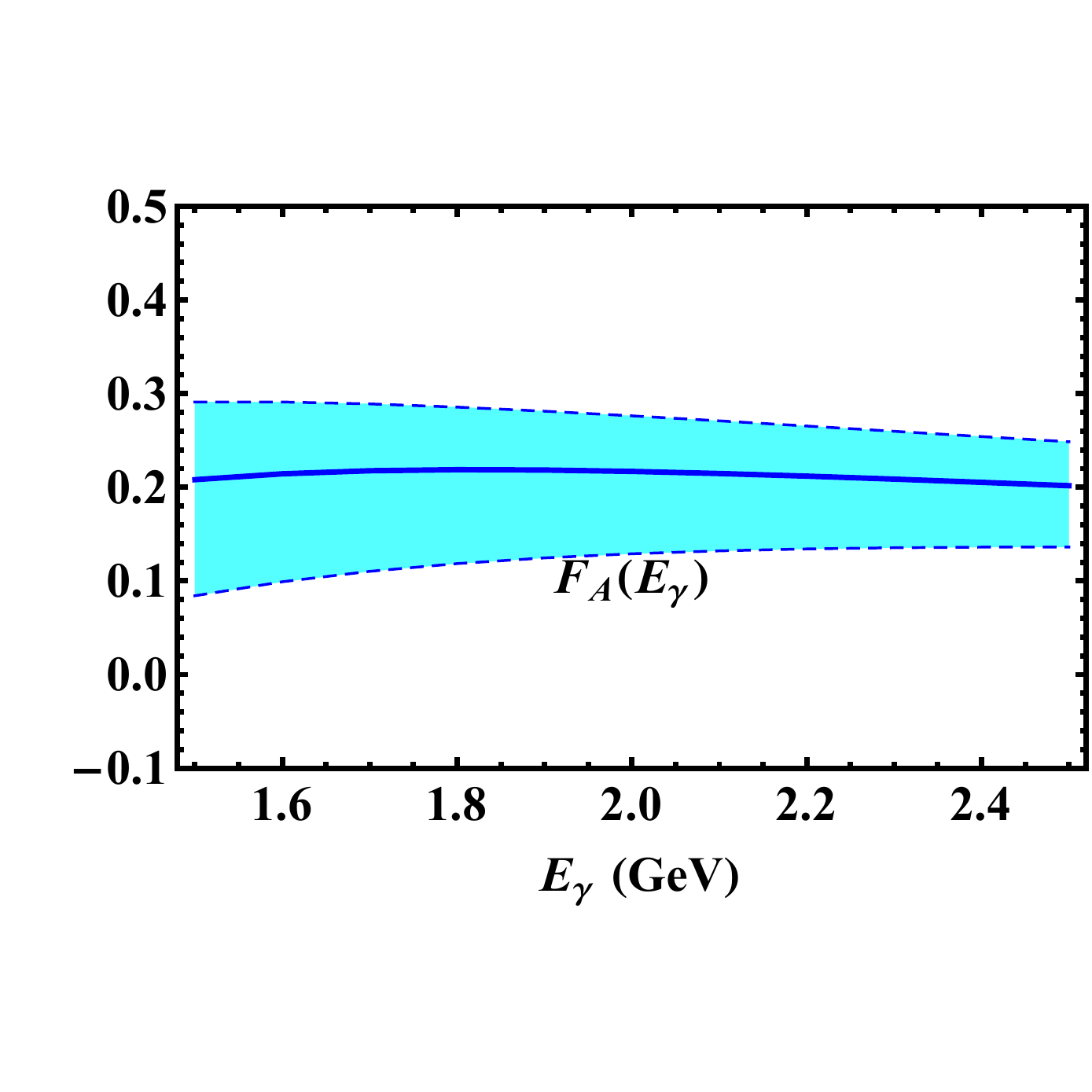}
\includegraphics[clip,width=0.48\textwidth]{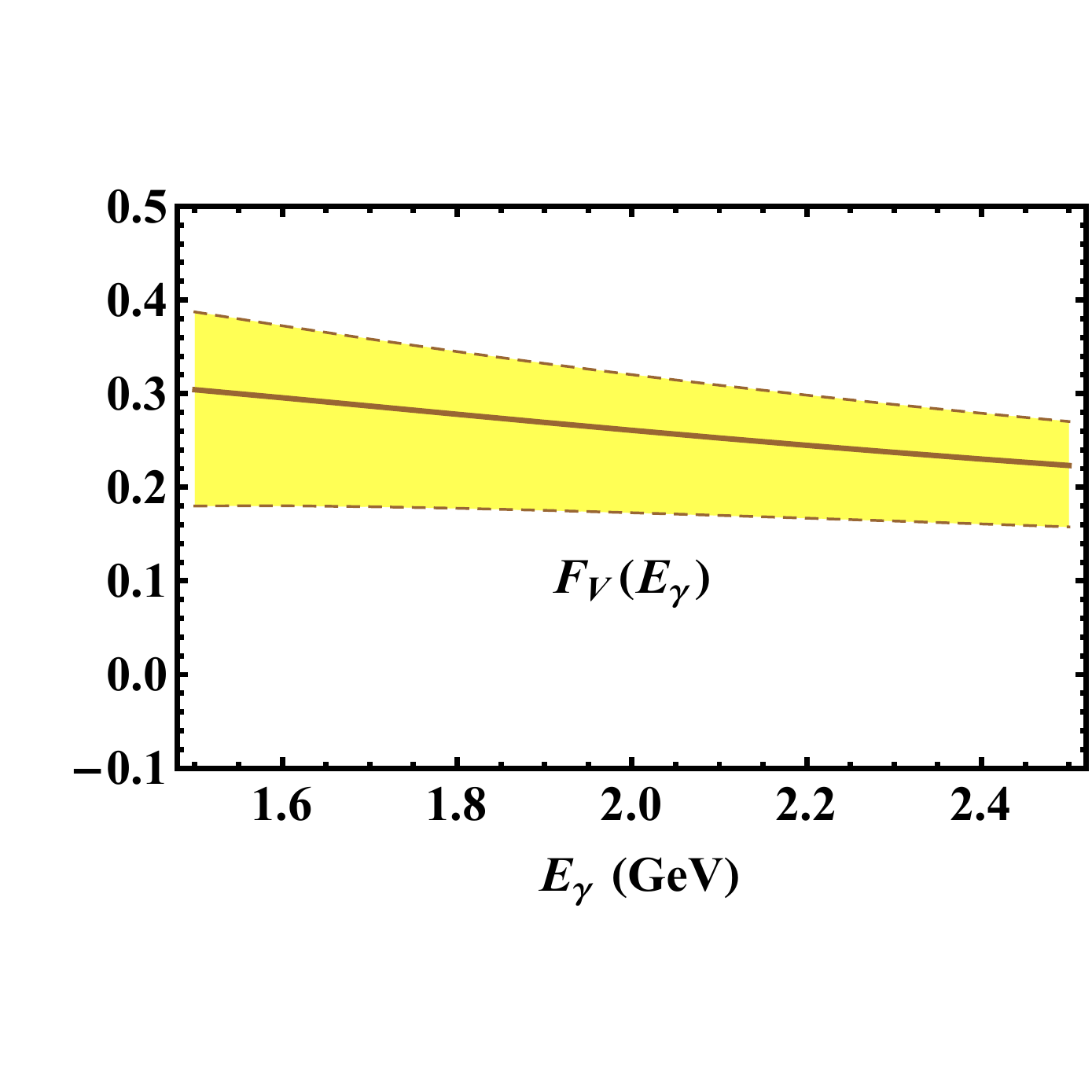}
\caption{The form factors $F_{V,A}$ with NLP contributions.
The uncertainties arise from varying parameters, including $\lambda_B$.
} \label{fig:ff}
\end{figure}

\begin{figure}
\centering
\includegraphics[clip,width=0.6\textwidth]{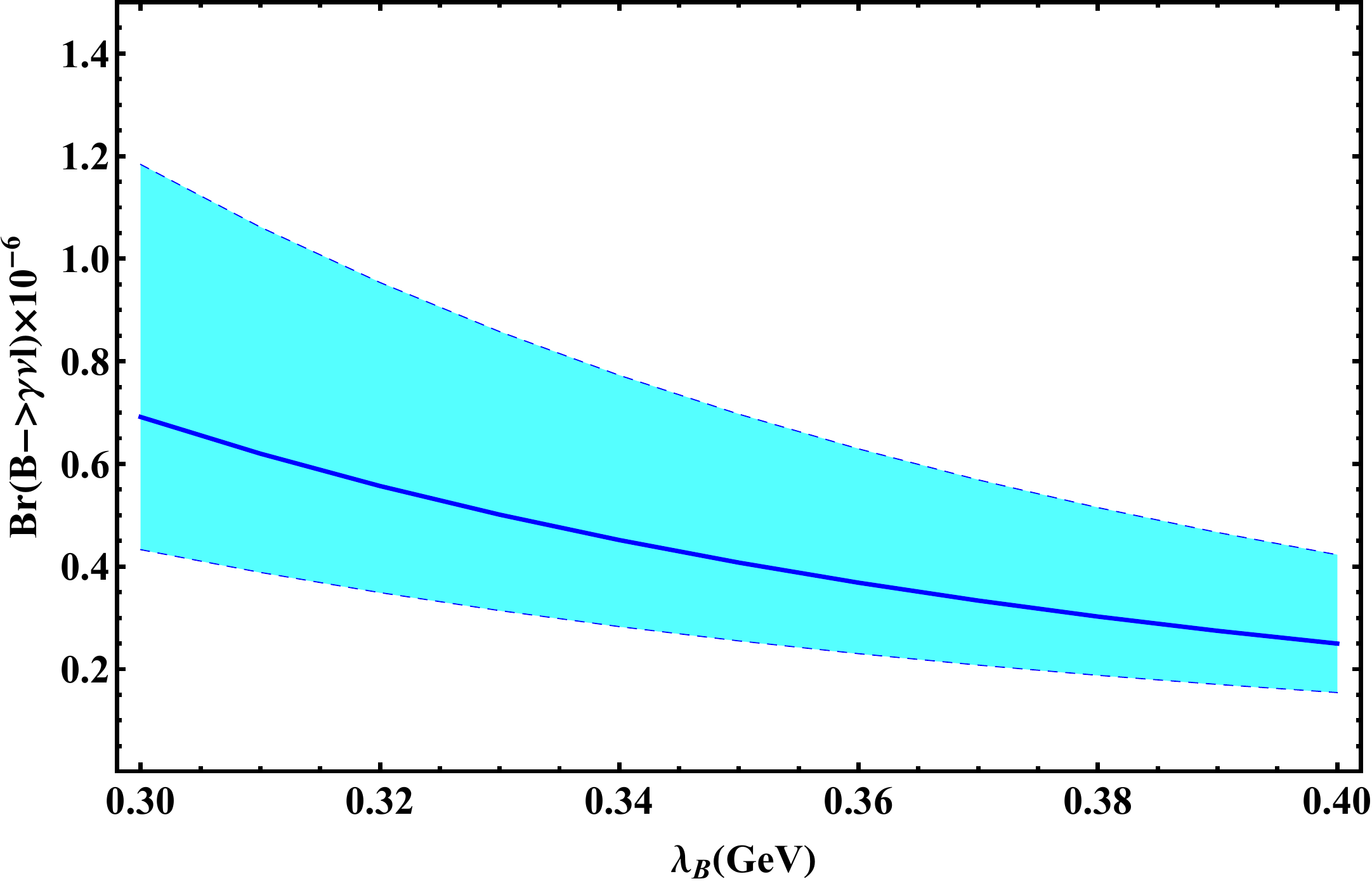}
\caption{The dependence of  ${\rm Br}(B\to \gamma\nu\ell)$ on $\lambda_B$ for the photon energy $E_{\gamma}>1.5~{\rm GeV}$.} \label{fig:br}
\end{figure}

The form factors including both the LP contribution and NLP corrections are plotted in Fig. \ref{fig:ff}, where for LP contribution, we adopt the same result as in \cite{Galda:2020epp}. Compared with the LP result in Fig. \ref{fig:fLL}, the power corrections significantly decrease the form factors, and the symmetry breaking effect from the NLP local contribution is also sizable. The uncertainties are denoted by the band in Fig. \ref{fig:ff}. To obtain the uncertainties, we consider various sources, including the following: the decay constant $f_B=0.192\pm0.0043 ~\mathrm{GeV}$; the first inverse moment $\lambda_B=0.35\pm0.05  ~\mathrm{GeV}$; the hard scale $\mu_h$ and hard-collinear scale $\mu_j$ (the same as in \cite{Galda:2020epp}); the models of the leading twist $B$-meson LCDA in Tab. (\ref{tab of parameters for B meson DAs}); the parameters $\bar \Lambda=0.48\pm 0.10 ~\mathrm{GeV} $, $\lambda_E^2\in[0.027 ~\mathrm{GeV}^2,0.088 ~\mathrm{GeV}^2]$, $\lambda_H^2\in[0.045 ~\mathrm{GeV}^2,0.222 ~\mathrm{GeV}^2]$; the Borel mass $M^2=1.25\pm0.25 ~\mathrm{GeV}^2$; and the threshold mass $s_0=1.5\pm 0.1 ~\mathrm{GeV}^2$ in the soft contribution.  The $\lambda_B$ parameter gives rise to the most important uncertainty as the LP result is inversely proportional to it.
Having the form factors at hand, the differential decay width is expressed as
\begin{eqnarray}
{d\Gamma\over d E_\gamma}={\alpha^2_{em}G_F^2|V_{ub}|^2\over 6\pi^2}m_BE_\gamma^3\left(1-{2E_\gamma\over m_B}\right)\left(|F_V|^2+\left|F_A+{Q_\ell f_B\over E_\gamma}\right|^2\right).
\end{eqnarray}
To guarantee the reliability of our calculation, we cut the photon energy at $E_\gamma >1.5  ~\mathrm{GeV}$.  We integrate over the differential decay width in the interval $[1.5 ~\mathrm{GeV}, m_B/2]$ and then multiply it by the lifetime of the $B$ meson to obtain the branching ratio ${\rm Br}(E_\gamma>1.5 ~\mathrm{GeV})$. If we fix $\lambda_B=0.35 ~\mathrm{GeV}$, the branching ratio reads
\begin{eqnarray}
{\rm Br}(B\to \gamma\nu\ell)=0.40^{+0.14}_{-0.24}\times 10^{-6},
\end{eqnarray}
 where the uncertainties come from the same source as that for the form factors (except for $\lambda_B$). The dependence of the branching ratio on $\lambda_B$ is presented in Fig. \ref{fig:br}, where the parameter $\lambda_B$ varies in the interval $[0.3~{\rm GeV},0.4~{\rm GeV}]$. We can see that the large uncertainty prevents us from precisely determinating of the parameter
$\lambda_B$.  It is thus important to reduce the uncertainty of the parameters, especially the uncertainty of the $B$ meson LCDA, and to obtain a more precise prediction of the power suppressed contribution.

\section{Summary}
The radiative leptonic decay mode $B \to \gamma\nu \ell$ is interesting both theoretically and experimentally. It plays an irreplaceable role in the determination of parameters of the $B$-meson LCDA. The factorization-scale dependence of the $B$-meson LCDA is governed by the LN evolution equation, which is an integro-differential equation and is not easily solved. We summarized the existing method of solving the LN evolution equation, for both the one-loop and two-loop anomalous dimensions.
We then derived a factorization formula with NLL resummation for the form factors appearing at leading power in the Mellin space, which is equivalent to the one obtained in \cite{Galda:2020epp} but written in a more compact form. The power corrections to the $B \to \gamma\nu \ell$ are sizeable, and much effort has been put into investigating the NLP contributions. In this paper, we included the power suppressed local contributions, the factorizable non-local contributions (which are suppressed by $1/E_\gamma$ and $1/m_b$), and the soft contributions.

In the numerical analysis, we found that the NLL-resummation effect significantly decreases the leading-power form factors, and that the NNLO correction brings approximately $5\%$ additional  reduction.  The NLP  contributions are combined with the leading-power NNLO contributions and also  manifestly decrease the form factors. We also calculated the integrated branching fractions of the  $B \to \gamma\nu \ell$ decay. The large  uncertainty from various sources makes it difficult to determine the parameter $\lambda_B$ and other logarithmic moments accurately. In future work, we will consider the NLP corrections more systematically in the framework of the SCET and  in the hope of reducing the theoretical uncertainty.

\subsection*{Acknowledgements}

Y.B.W is supported in part by the National Natural Science Foundation
of China (NSFC) with Grant No. 11675082 and 11735010.

\end{document}